\documentstyle[sprocl]{article}

\def\be{\begin{equation}}
\def\ee{\end{equation}}
\def\bea{\begin{eqnarray}}
\def\eea{\end{eqnarray}}

\def\beqa{\begin{eqnarray}}
\def\eqa{\end{eqnarray}}

\newcommand{\bb}{\mbox{$\underline {b}$}}

\newcommand{\wb}{\mbox{$\underline {w}$}}

\newcommand{\de}{\mbox{$\frac{1}{2}$}}

\def\beq{\begin{equation}}
\def\eq{\end{equation}}

\begin{document}
\input epsf
\begin{titlepage}
\vspace*{-1.5cm}

\begin{center}
\baselineskip=13pt

{\Large \bf The equivalence between the color dipole model and the BFKL 
Pomeron at leading order.\\}
\vskip1.5cm
{\Large Samuel Wallon}
\vskip1.5cm
{\it II. Institut f\"ur Theoretische Physik, Universit\"at Hamburg \\ 
 Hamburg, Germany\\ {\rm and} \\
Division de Physique Th\'eorique,\footnote{Unit\'e
 de Recherche des Universit\'es Paris 11 et Paris 6 Associ\'ee au CNRS} 
Institut de Physique Nucl\'eaire d'Orsay \\ 
91406 Orsay, France \\
{\rm and} \\
LPTPE, Universit\'e P. \& M. Curie, 4 Place Jussieu \\
 75252 Paris Cedex 05, 
France}
\end{center}
\vspace*{.8cm}
\begin{quotation}
\centerline{\bf Abstract}
\vspace*{.4cm}
We compute the onium-onium scattering amplitude at fixed impact parameter
in the framework of the perturbative QCD dipole model. Relying on the conformal
properties of the dipole cascade and of the elementary dipole-dipole 
scattering amplitude, we obtain an exact result for this onium-onium
 scattering amplitude, which is proved to be identical to the BFKL result,
 and which 
is frame invariant. The asymptotic
expression for this amplitude 
at fixed impact parameter agrees with previous numerical simulations.
\vskip 1.0cm
\noindent
Talk given at the Madrid workshop on low x physics.

\noindent
June 18-21, 1997, Miraflores de la Sierra, Spain.
\end{quotation}

\vspace{1.1cm}
\noindent
IPNO/TH 97-36
\end{titlepage}
\mbox{}
\thispagestyle{empty}
\newpage
\setcounter{equation}{0}
\setcounter{page}{1}



\section{Introduction}
\label{introduction}

The small-$x_{bj}$ HERA data \cite{h1zeus} have focused attention on the
 Balitsky-Fadin-Kuraev-Lipatov (BFKL) Pomeron.\cite{lip} In this
 small-$x_{bj}$
 regime, the soft singularities lead to 
  expansion for the proton structure functions of the type 
$\sum_{p \geq n} \alpha_s^p (\ln 1/x_{bj})^n.$
The leading behaviour $(LL1/x_{bj}),$ corresponding to $p=n,$
 has been known for many years at fixed $\alpha_s.$\cite{lip}
This calculation was performed computing order by order the contribution of
the relevant Feynman diagrams up to $g^8,$ and then reconstructing the full 
amplitude by imposing unitarity order by order. 
This perturbative QCD hard Pomeron describes the behaviour of hadronic
 scattering
 amplitudes at very high energy $s$ and fixed momentum transfer $t$.
 It is a  bound state of two reggeized gluons in $t$-channel, which predicts
 a behaviour of the amplitude of the type
\begin{equation}
\label{ABFKL}
A_{BFKL} \propto s^{\alpha_P} \ \ \ {\rm with} \ \ \ 
\alpha_P = 1 + \frac{\alpha_S N_c}{\pi} 4 \ln 2 > 1.
\eq
This behaviour violates
 the Froissart bound at very high $s$
\beq
\label{bornefroissart}
\sigma_{tot} \leq c \, \ln^2 s.
\eq
This violation is directly related to the unitarization problem of QCD, which 
is one of the main problems to be solved in the theory of strong interaction.
The Generalized Leading 
Logarithmic Approximation,\cite{BJKP} where one takes into 
account the exchange of any fixed number of reggeized gluons in $t$-channel, 
has been proposed in order to solve this problem. It  is proved to be
 equivalent to the integrable non-compact Heisenberg XXX spin chain 
\cite{lipxxxfk} in
 the
 multicolor limit of QCD. The solution of this integrable model is still an 
open 
problem.\cite{resolution}

In the model recently developed by Mueller et 
al \cite{mueller94,muellerpatel,muellerunitarite,muellerchen} 
and separately by Nikolaev et al,\cite{nikital}
 in order to control the perturbative approach, one deals
 with 
onia, which are heavy quark-antiquark bound states, so that their transverse
 size 
naturally 
provides an infra-red cut-off. The relevant degrees of freedom at high energy
 are
then
made of color dipoles. In the multicolor limit, the dominant topology in color
space is the  cylinder, and these color dipoles produce a classical cascade,
 which reveals a Pomeron type dynamics. 
Combining this dipole model or the BFKL Pomeron with
 $k_T$-factorization,\cite{catanicollinslevin} it is possible to describe 
deep inelastic 
$e^\pm-p$ scattering at HERA in the small-$x_{bj}$ 
regime.\cite{nprnprwwal7,agkms}  In the more general case of onium-onium
scattering, the BFKL approximation corresponds to the 
exchange of one pair of gluons between two excited dipoles, each one being
extracted
 from one of the two onia.   
The unitarization problem can also be studied in this dipole model, which 
gives an $s-$channel picture of the process rather than a $t-$channel
 description as in the BFKL 
approach.\cite{muellerunitarite,salam,muellersalam,kmw} 
Unitarity implies that 
 the probability of any event
 cannot exceed 1, that is 
\beq
|S(b)| \leq 1.
\eq
It is thus important to evaluate precisely the scattering amplitude at fixed
impact parameter, in order to see how the theory unitarizes. 
The equivalence between the BFKL and the dipole approach has been proved 
formally, by making a comparison between the Feynman diagrams involved in the
 perturbative Regge approach and the time ordered graphs of the dipole 
model.\cite{muellerchen} However the corresponding analytic expression of the
 amplitudes were only proved to be equal asymptotically for very large 
relative rapidity $Y.$\cite{muellerpatel,muellerunitarite} In this
 contribution, we prove the exact
equivalence of the two amplitudes at LLA at fixed impact parameter,
relying on the global conformal invariance of the
 model. We also obtain the exact distribution
of the dipole in transverse impact parameter space when $Y \gg 1,$
which is compatible with previous numerical and analytical estimations.
An enlarged version of this paper can be found in Ref.\cite{nw}

\section{Onium-onium cross-section at fixed impact parameter}
\label{onium2}
Consider onium-onium scattering in the leading logarithmic
 approximation. Due to factorization of the soft gluonic part of the onia
wave function, one can use a parton type formulation for calculating
the onium-onium
scattering amplitude $A$ at fixed impact parameter. It involves the number of
 dipoles in each onium and the elementary cross-section of two such dipoles. 
 For a relative rapidity $Y$ 
and impact parameter $b$, $A(Y,\underline{b})$ can be expressed as
\beq
\label{Ab}
A(Y,\underline{b}) = -i \int d^2 \underline{x}_1
 \, d^2 \underline{x}_2 \int^1_0 dz_1 \, dz_2 \, 
 \Phi(\underline{x}_1,z_1) \,  \Phi(\underline{x}_2,z_2) \,
 F(\underline{x}_1,\underline{x}_2,\tilde{Y},\underline{b}).
\eq
$\Phi(\underline{x}_i,z_i)$ is the square of the heavy quark-antiquark part of 
the onium
 wavefunction, $\underline{x}_i$ being the transverse size of the 
quark-antiquark pair
 and $z_i$ the longitudinal momentum fraction of the antiquark. The momentum
$p_1^+$ and $p_2^-$ of the two onia are supposed to be large, with
$\underline{p}_1 = \underline{p}_2 = 0.$ $Y$
is related to $\tilde{Y}$ by $\tilde{Y} = Y + \ln z_1 z_2,$ due to the fact
 that the 
{\it perturbative} dipole cascade originates from the quark-antiquark pairs. 
The 
distributions $\Phi(\underline{x}_i,z_i)$ of these pairs cannot be computed
 perturbatively, and goes far beyond the purpose of the present approach.
In the leading logarithm
approximation (noted $F^{(1)}$), the scattering is due to the exchange
 of a single pair of gluons between the two dipoles extracted from the left
 and right moving onia.
The process is illustrated 
\begin{figure}[htb]
\begin{picture}(500,165)(0,0)
\put(-95,96){\epsfysize=4.8cm{\centerline{\epsfbox{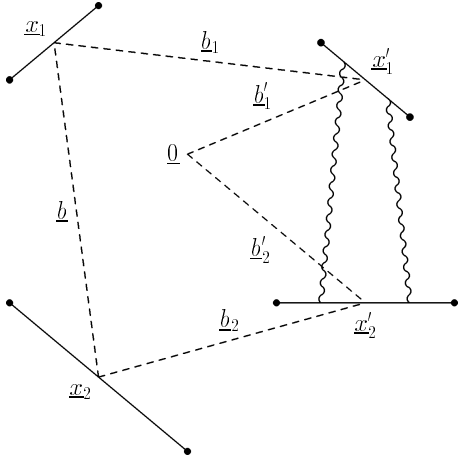}}}}
\end{picture}
\caption{Onium-Onium scattering at leading order.}
\label{diffusion}
\end{figure}
in 
 fig. \ref{diffusion}.  We denote $\underline{x}_{a1}$ 
($\underline{x}_{b1}$) the
 transverse
 coordinate of
 the heavy quark
 (antiquark) making up the right moving onium and
 $\underline{x}_{a2}$ ($\underline{x}_{b2}$) the coordinates of the
 corresponding
quark (antiquark) making the left moving onium. These onia of transverse sizes
 $\underline{x}_1 = \underline{x}_{a1} - \underline{x}_{b1}$ and 
$\underline{x}_2 = \underline{x}_{a2} - \underline{x}_{b2}$ 
 scatter through  the exchange of a pair of
gluons between two elementary dipoles, respectively of transverse sizes
 $\underline{x}'_1$ and
 $\underline{x}'_2$, located at  $\underline{b}'_1$ and $\underline{b}'_2$
 with respect to the
 reference
 point $\underline{0}$ (which is arbitrary due to translation invariance).
 These two
 elementary dipoles are produced by the two heavy quark-antiquark pairs at a
 distance
 $\underline{b}_1$ and $\underline{b}_2$ from their center of mass. 
$F^{(1)}$ thus reads
\beqa
\label{F1b1}
&& \hspace{-1cm} F^{(1)}(\underline{x}_1,\underline{x}_2,\tilde{Y},
\underline{b}) =
 -\de \int \frac{d^2 \underline{x}'_1}
{2 \pi {x'_1}^2} \frac{d^2 \underline{x}'_2}{2 \pi {x'_2}^2}
 d^2 \underline{b}_1 
\,d^2 \underline{b}_2 \, \delta^2
(\underline{b}_1 - \underline{b}_2 - \underline{b}'_1 + \underline{b}'_2 - 
\underline{b}) \nonumber \\
&&\hspace{-.5cm}\times \, d^2 (\underline{b}'_2 - \underline{b}'_1)\,
 n(\underline{x}_1,\underline{x}'_1,\tilde{Y}_1,\underline{b}_1) \,
 n(\underline{x}_2,\underline{x}'_2,\tilde{Y}_2,\underline{b}_2)
 \, \sigma_{DD}(
\underline{x}'_1,\underline{x}'_2,\underline{b}'_1 - \underline{b}'_2) .
\eqa
The rapidities $\tilde{Y}_1$ and $\tilde{Y}_2$ are such that 
$\tilde{Y}=\tilde{Y}_1 +
 \tilde{Y}_2.$
Eq. (\ref{F1b1}) involves the elementary cross-section 
$\sigma_{DD}(\underline{x}'_1,
\underline{x}'_2,\underline{b}'_1 - \underline{b}'_2)$
of two dipoles
of transverse sizes $\underline{x}'_1$ and $\underline{x}'_2$,
whose centers are located at $\underline{b}'_1$ and $\underline{b}'_2,$
 which has been evaluated in Refs.\cite{muellerpatel} and \cite{nw},
 and the number density $n(\underline{x},\underline{x}',\tilde{Y},
\underline{b})$ of dipole of transverse size $\underline{x}'$ at a 
transverse distance $\underline{b}$ from the center of the quark-antiquark 
pair of transverse size
$\underline{x},$ at relative rapidity $\tilde{Y}$ (see Ref.\cite{nw} 
for precise definitions). 

In order to compute exactly the amplitude $F^{(1)}$  given by Eq.
(\ref{F1b1}),
we rely on the global conformal invariance of the process, which enable us to 
expand both the dipole distribution
 and the 
dipole-dipole cross-section on a suitable basis of three points holomorphic 
and antiholomorphic correlation functions.\cite{lipatov86,polyakov} 
 Introducing the complex 
coordinates $\rho = \rho_x + i\rho_y$ and $\rho^* = \rho_x - i\rho_y$
in the two-dimensional transverse space, 
the complete set of eigenfunctions $E^{n,\nu}$ of the dipole emission kernel 
is
\beq
\label{defE}
E^{n, \nu}(\underline{\rho}_{10},\underline{\rho}_{20}) = (-1)^n\left(\frac{
\rho_{12}}{\rho_{10} \rho_{20}} \right)^h \left(\frac{\rho^*_{12}}{\rho^*_{10}
 \rho^*_{20}} \right)^{\bar h},
\eq
$h = \frac{1-n}{2} + i \nu$ and $\bar h = \frac{1+n}{2} + i \nu$
being the corresponding conformal weights, with $n$ integer and $\nu$ real.

We get rid of the longitudinal degrees of freedom by using a Mellin transform 
with respect to $\tilde{Y}$, namely
\beq
\label{mellinomega}
n(x,x',\tilde{Y},b) = \int \frac{d \omega}{2 i \pi} e^{\omega \tilde{Y}}
 n_{\omega}(x,x',b).
\eq
Expanding the dipole distribution on the conformal basis, one writes
\beqa
\label{decn}
&&\hspace{-.3cm}n_{\omega}(x,x',b) = \sum_{n=-\infty}^{n=+\infty} 8 \int 
\frac{d \nu}{(2 \pi)^3}
\frac{d^2 w}{x'^2} \left(\nu^2 + \frac{n^2}{4} \right) n_{\{\nu,n\}\omega}
 \nonumber \\
&&\hspace{.2cm}\times \,  E^{n,\nu}\left(\bb + \frac{\underline{x}'}{2} - 
\wb,\bb-
\frac{\underline{x}'}{2} - \wb \right) E^{n,\nu*}\left(\frac{\underline{x}}
{2} -\wb,-\frac{\underline{x}}{2} - \wb \right).
\eqa
The longitudinal dynamics gives rise to the term $n_{\{\nu,n\}\omega},$
which reads \cite{mueller94}
\beq
\label{distn1}
n_{\{\nu,n\}\omega}= \frac{2}{\omega - \frac{2 \alpha_S N_c}{\pi} \chi(n,\nu)},
\eq
where
\beq
\label{chinnu}
\chi(n,\nu) = \psi(1) - {\rm Re} \, 
\psi\left(\frac{|n|+1
}{2} + i \nu\right).
\eq
Expanding the dipole-dipole 
cross-section on the conformal basis yields \cite{nw}
\beqa
\label{rappelsigmaDD}
&&\hspace{-.7cm} \sigma_{DD}(\underline{x}'_1,
\underline{x}'_2,\underline{b}'_1-\underline{b}'_2) \!=\! 
\frac{2 \alpha_s^2}{(2 \pi)^2}\! \! \sum_{n=-\infty}^{+\infty}\!
 \int_{-\infty}^{+\infty}\! \!
 d \nu \! \int \! d^2 \underline{w} 
 \frac{\left(\nu^2 + \frac{n^2}{4}\right)\left(1 +
 (-1)^n\right)}{\left(\!\nu^2 + \left(\frac{n-1}{2} \right)^2 \right) 
\! \left(\! \nu^2 + 
\left ( \frac{n+1}{2} \right)^2 \right)} \nonumber \\
&&\hspace{-.8cm}\times \, E^{n,\nu*}\left(\underline{b}'_1 +
 \frac{\underline{x}'_1}{2} - 
\underline{w},\underline{b}_1 - \frac{\underline{x}'_1}{2} -
 \underline{w}\right) \!
 E^{n,\nu}\left(\underline{b}'_2 + \frac{\underline{x}'_2}{2} - \underline{w},
\underline{b}'_2 - \frac{\underline{x}'_2}{2} - \underline{w}\right).
\eqa
The full expression for $F^{(1)}$ now reads,
combining Eqs. (\ref{F1b1},  \ref{mellinomega}, \ref{decn}, \ref{rappelsigmaDD}),
\beqa
\label{calculF1}
&&\hspace{-.7cm} F^{(1)}(\underline{x}_1,\underline{x}_2,\tilde{Y},
\underline{b}) \!=\! -\frac{\alpha_s^2 (16)^2}{(2 \pi)^2}
\! \! \sum_{n_1 = -\infty}^{+\infty} \sum_{n_2 = -\infty}^{+\infty} \sum_{n =
 -\infty}^{+\infty} \int^{\infty}_{-\infty} \! \frac{d \nu_1}{(2 \pi)^3}
 \int^{\infty}_{-\infty} \!
 \frac{d \nu_2}{(2 \pi)^3}\! \int^{\infty}_{-\infty}\! d \nu \nonumber\\
&&\hspace{-.7cm} \times  
 \frac{\left(\nu^2 + \frac{n^2}{4}\right)
\left(1 + (-1)^n \right)}{\left( \nu^2 + \left(\frac{n-1}{2} \right)^2 \right)  \left(
 \nu^2 +
 \left ( \frac{n+1}{2} \right)^2 \right)}  \exp
 \left(\frac{2 \alpha_s N_c}{\pi}(\chi(n_1,\nu_1)\tilde{Y}_1
 +\chi(n_2,\nu_2)\tilde{Y}_2)\right)
\nonumber\\
&&\hspace{-.7cm} \times \left(\nu_1^2 + \frac{n_1^2}{4} \right)   
\left(\nu_2^2 + \frac{n_2^2}{4}
 \right)
\int
 \frac{d^2\underline{w}_1}{{x'_1}^2} \,
 \frac{d^2\underline{w}_2}{{x'_2}^2} \, d^2 \underline{w}  \int \frac{d^2
 \underline{x}'_1}{2 \pi {x'_1}^2}  \, \frac{d^2 \underline{x}'_2}
{2 \pi {x'_2}^2} \, d^2
 \underline{b}_1 \, d^2 \underline{b}_2 \, d^2 \underline{b}_{int} 
 \nonumber \\
&&\hspace{-.7cm}\times \,  \delta^2(\underline{b}_1 - \underline{b}_2 +
 \underline{b}_{int} - \underline{b}) \, E^{n_1,\nu_1}\left(\underline{b}_1 +
 \frac{\underline{x}'_1}{2} - \underline{w}_1,\underline{b}_1 -
 \frac{\underline{x}'_1}{2} -
 \underline{w}_1\right) \, \nonumber \\
&&\hspace{-.7cm}\times \, E^{n_1,\nu_1*}\left(\frac{\underline{x}_1}{2} -
 \underline{w}_1,-
 \frac{\underline{x}_1}{2} - \underline{w}_1\right)\,
 E^{n_2,\nu_2*}\left(\underline{b}_2 +
\frac{\underline{x}'_2}{2} - \underline{w}_2,\underline{b}_2 -
 \frac{\underline{x}'_2}{2} -
 \underline{w}_2 \right) \, \nonumber \\
&&\hspace{-.7cm}\times \, E^{n_2,\nu_2}\left(\frac{\underline{x}_2}{2} -
 \underline{w}_2, -
 \frac{\underline{x}_2}{2} - \underline{w}_2\right)\,
 E^{n,\nu*}\left(\underline{b}_1 +
 \frac{\underline{x}'_1}{2} - \underline{w},\underline{b}_1 -
 \frac{\underline{x}'_1}{2} - \underline{w}\right)  \nonumber \\
&&\hspace{-.7cm}\times \, E^{n,\nu}\left(\underline{b}_{int} + \underline{b}_1
 +
\frac{\underline{x}'_2}
{2} -
 \underline{w}_1,\underline{b}_{int} + \underline{b}_1 -
 \frac{\underline{x}'_2}
{2} - 
\underline{w}_1\right) \,  ,
\eqa
where we have  set 
$\underline{b}_{int} =\underline{b}'_2 - \underline{b}'_1.$
The quantum numbers $n_1,\nu_1$ and $n_2,\nu_2$
correspond respectively to the distributions
 $n(\underline{x}_1,\underline{x}'_1,
\tilde{Y}_1,\underline{b}_1)$ and $n(\underline{x}_2,\underline{x}'_2,
\tilde{Y}_2,\underline{b}_2)$.
The integration with respect to $\underline{b}_2$ is done through the
 delta
 distribution.  
 Using the
 orthonormalization condition for the functions
 $E^{n,\nu},$\cite{lipatov86,nw}
\beqa
\label{ortho}
&&\hspace{-1cm} \int \frac{d^2 \underline{\rho_1} \,
 d^2 \underline{\rho_2}}{|\rho_{12}|^4} E^{n,\nu}(\rho_{10},\rho_{20}) \,
 E^{m,\mu*}(\rho_{10'},\rho_{20'})= a_{n,\nu}
 \delta_{n,m} \, \delta(\nu -\mu) \, \delta^2(\rho_{00'}) \nonumber \\
&&+ (-1)^n \, b_{n,\nu}|\rho_{00'}|^{-2-4 i \nu} (\rho_{00'}/\rho_{00'}^*)^n
 \delta_{n,-m} \,
 \delta(\nu+\mu)\, ,
\eqa
the integration over  $d^2 \underline{b}_1 \, d^2 \underline{b}_{int} \, d^2
 \underline{x}'_1 \, d^2 \underline{x}'_2$ gives four
terms with factors $a_{n_1,\nu_1},$ $a_{n_2,\nu_2},$
  $b_{n_1,\nu_1}$ and $b_{-n_2,-\nu_2}.$
 $E^{n,\nu}$ and $E^{n,\nu*}$ being related by
 the relation
 \cite{lipatov86,nw}
\beq
\label{lien}
\hspace{-.2cm} E^{n,\nu*}(\underline{\rho}_{10},\underline{\rho}_{20}) \!= 
\! \frac{b^*_{n,\nu}}
{a_{n,\nu}} \! \int \! d^2 \underline{\rho}_{0'}
 E^{n,\nu}(\underline{\rho}_{10'},\underline{\rho}_{20'}) |\rho_{00'}|^{-2 + 4
 i
 \nu}\! \left(\frac{\rho^*_{0'0}}
{\rho_{0'0}}\right)^n \! (-1)^n,
\eq
the contribution of these four terms are in fact identical.
The integration with respect to $\underline{w}_1$ and $\underline{w}_2$ can
 then be
 performed, and using 
 $\tilde{Y}=\tilde{Y}_1 + \tilde{Y}_2,$ one finally gets
\beqa
\label{calculF3}
&&\hspace{-1.4cm} F^{(1)}(x_1,x_2,\tilde{Y},b) = -\frac{\alpha_s^2}{(2 \pi)^2}
\sum_{n = -\infty}^{+\infty} \int^{\infty}_{-\infty} d \nu   \int d^2 
\underline{w} 
 \left(\nu^2 + \frac{n^2}{4}\right) \nonumber \\
&&\hspace{-1.2cm} \times \, \frac{1 + (-1)^n}{\left( \nu^2 +
 \left(\frac{n-1}{2} \right)^2 \right) \left( \nu^2 + \left ( \frac{n+1}{2} 
\right)^2
 \right)}  \exp \left(\frac{2 \alpha_s N_c}{\pi}\chi(n,\nu)\tilde{Y}
\right) \nonumber\\
&&\hspace{-1.2cm} \times \,
 E^{n,\nu*}\left(\frac{\underline{x}_1}{2} - \underline{w},- 
\frac{\underline{x}_1}{2}
 - \underline{w}\right) \,E^{n,\nu}\left(\frac{\underline{x}_2}{2} - 
\underline{w}
 +
 \underline{b}, - \frac{\underline{x}_2}{2} - \underline{w} +\underline{b}
 \right).
\eqa
This result, obtained without any 
approximation,
 is clearly independent of the choice of the reference frame, since
 the
result only depends on the total rapidity $\tilde{Y}.$ 
It also proves the equivalence between the dipole and the
 BFKL
approaches at leading order. Indeed, taking into account form factors
when coupling  the $t$-channel bound state of reggeized gluons to the external
 quark-antiquark pairs 
 and the difference of 
definition
of amplitudes, one should fulfil
\beq
\label{equivdipoleBFKL}
F^{(1)}(\underline{x}_1,\underline{x}_2,\tilde{Y} \!,\underline{b}) \!= \!
 -\frac{\alpha_s^2}{(2 \pi)^2}  \!\int  \!
 \frac{d \omega}{2 \pi i} \,e^{\omega \tilde{Y}} 
 [f_{\omega}(\underline{x}_{a1},\underline{x}_{b1},\underline{x}_{a2},
\underline{x}_{b2}) + (a_1 \! \leftrightarrow  \! b_1)],
\eq
where $f_{\omega}$ is the BFKL expression defined by equation (26) of
 Ref. \cite{lipatov86}
\beqa
\label{fomega}
&&\hspace{-1.3cm} f_{\omega}(\underline{x}_{a1},\underline{x}_{b1},
\underline{x}_{a2},
\underline{x}_{b2}) = \sum_{n=-\infty}^{+\infty} \int_{-\infty}^{+\infty}
 \int d^2\underline{x}_0 
 \frac{\nu^2 + \frac{n^2}{4}}{\left( \nu^2 +
 \left(\frac{n-1}{2} \right)^2 \right) \left( \nu^2 + \left ( \frac{n+1}{2} 
\right)^2
 \right)} \nonumber\\
&&\hspace{-1.3cm}\times \, \frac{1}{\omega - \frac{2 \alpha_s N_c}{\pi}\chi(n,
\nu)} 
 E^{n,\nu*}\left(\underline{x}_{a2} - \underline{x}_0,
 \underline{x}_{b2} - \underline{x}_0\right)
 E^{n,\nu}\left(\underline{x}_{a1} - \underline{x}_0,
 \underline{x}_{b1} - \underline{x}_0 \right).
\eqa
 A straighforward check shows that Eq. (\ref{equivdipoleBFKL}) is
 satisfied by $F^{(1)}$ and  $f_{\omega}$.
This equivalence can also be proved by comparing the real 
and virtual graphs in covariant (BFKL) and light-cone (dipole) quantization.
 The result is that
the sum of real and virtual contributions is identical in both case, although
each of these terms differs. Thus, this result is true only for
 inclusive quantities.\cite{muellerchen} 

In the asymptotic regime the term corresponding to 
$n=0$ dominates, and integrating over $\nu$ by a saddle point method
(the dominant contribution comes from $\nu \sim 0$ due to 
$\chi(0,\nu)$ given by Eq. (\ref{chinnu})), one gets from Eq. (\ref{calculF3})
\beqa
\label{resFexact}
&&\hspace{-1.5cm} F^{(1)}(\underline{x}_1,\underline{x}_2,\tilde{Y},
\underline{b}) \nonumber \\
&&\hspace{-1.5cm} \simeq \! - \pi \alpha_s^2\frac{x_1 \, x_2}{b^2}
 \frac{\ln (16 \, b^2/x_1 x_2)}{\left(\frac{7}{2} \alpha_s N_c  \zeta(3)
 \tilde{Y}\right)^{3/2}} \exp \left \{\displaystyle
 \frac{4 \alpha_s N_c}{\pi}\ln 2 \, \tilde{Y} - \frac{\ln^2 (16\,
  b^2/x_1 x_2)}{\frac{14 \alpha_s N_c}
{\pi}
\zeta(3) \tilde{Y}} \right \} 
\eqa
in the domain 
\beq
\label{domaineF}
\frac{2 \alpha_s N_c}{\pi}7 \zeta(3) \tilde{Y} \gg 
\ln \frac{16 b^2}{x_1 x_2} \gg 1.
\eq
This result, which differs from Eq. (10) of Ref. 
\cite{muellerunitarite} by a factor $16$, is 
in agreement with numerical simulations.\cite{salamthesis}
The total cross-section $F^{(1)}_{tot}$ can be obtained by integrating the 
cross-section at 
fixed impact parameter.
However a direct calculation starting 
from the integrated dipole distribution is much easier \cite{nw}
and gives, in the asymptotic regime $Y \gg 1,$
\beqa
\label{Ftotal4}
F^{(1)}(\underline{x}_1,\underline{x}_2,\tilde{Y}) \simeq 
-2 \pi \, \alpha_s^2 \, x_1 \,x_2 
 \frac{\exp \left \{\displaystyle
 \frac{4 \alpha_s N_c}{\pi}\ln 2 \, \tilde{Y} 
-\displaystyle \frac{\ln^2 (x_1/ x_2)}{\frac{14 \alpha_s N_c}{\pi}\zeta(3) 
\tilde{Y}} \right \}}{\sqrt{\frac{7}{2} \alpha_s N_c \zeta(3) \tilde{Y}}}
\eqa
in agreement with formula (26) of Ref. \cite{muellerpatel}.
This result is valid in the range
\beq
\label{domaineF2}
\frac{2 \alpha_s N_c}{\pi}7 \zeta(3) \tilde{Y} \gg 
\left|\ln \frac{x_1}{x_2} \right| \gg 1.
\eq
Since we are interested here in the domain were the two onia have comparable
 transverse sizes, this condition is satisfied in a wide range in $\tilde{Y}$.
 However, in the case of $e^{\pm}-onium$
deep inelastic scattering, this provides some limitations to the approximation.
Indeed the two relevant scales, namely the transverse size of the onium (or
an effective
scale in the case of proton) and the inverse of the virtuality of the electron,
are then very different, leading to a shift of the saddle point in the 
imaginary axis towards the region $\nu =i/2$ where the approximation is not
 valid anymore. This domain is precisely the one were the DGLAP approximation
 becomes
competitive with the BFKL approximation.

Let us integrate the scattering amplitude $F^{(1)}(\underline{b})$ 
with respect to the impact parameter $\underline{b}$ in the domain 
(\ref{domaineF}) where formula (\ref{resFexact}) is valid 
(neglecting the fact that the upper bound is not infinite).
This gives
\beq
\label{integreFbexact}
F^{(1)}(\underline{x}_1,\underline{x}_2,\tilde{Y})
\simeq -2 \pi \, \alpha_s^2 \, x_1 \,x_2 
 \frac{\exp \left \{\displaystyle
 \frac{4 \alpha_s N_c}{\pi}\ln 2 \, \tilde{Y} \right \}}{\sqrt{\frac{7}{2} \alpha_s 
N_c 
 \zeta(3)
 \tilde{Y}}}  \exp
 \left \{-\displaystyle \frac{\ln^2 (x_1/ x_2)}{\frac{14 \alpha_s N_c}{\pi}
\zeta(3) \tilde{Y}} \right \} \,,
\eq
which is identical to Eq. (\ref{Ftotal4}). From the gaussian distribution
obtained in Eq. (\ref{integreFbexact}), it is clear, comparing with 
Eq. (\ref{resFexact}), that the total cross 
section at BFKL order is dominated by impact parameter configuration 
much larger than the transverse sizes of the two scattering onia,
corresponding to
\beq
\label{ordreb}
\ln \left(\frac{16 \, b^2}{x_1 x_2}\right) \sim \sqrt {\frac{14 \alpha_s N_c}{\pi}
\zeta(3) \tilde{Y}}.
\eq
Note that this dominant contribution is inside the domain (\ref{domaineF}).
These dominant configurations are  much more central than what was claimed in
 Ref.\cite{muellerunitarite} 
It confirms previous numerical simulations.\cite{salam} 
Thus, the calculation, based on perturbative 
QCD, is expected to remain valid for high values of $\tilde{Y}.$

Combining this dipole model with $k_T-$factorization, it is possible to get
a full description of $e^{\pm}-onium$ deep inelastic scattering at low 
$x_{bj}.$\cite{nw} 
The application of this analysis to the proton \cite{nprnprwwal7} requires
 some additional asumptions for the coupling of the dipole
cascade to the proton.
 It leads to a successful description of the HERA data, and
 provides a
 prediction for the gluon density and for $R= F_L/F_T$.
 This model can also be applied to diffractive 
physics.\cite{bp}
\section{Conclusion}
We have shown the exact equivalence between BFKL and dipole
approaches for the onium-onium cross-section at fixed impact parameter. This
 proof
relies on conformal properties of the dipole cascade and of the elementary 
dipole-dipole cross section. We have also obtained an asymptotic expression
 for the  onium-onium
 cross-section at fixed impact parameter, which agrees with previous 
numerical simulations. 
As it has been seen previously, the dipole model
can only be safely applied when the two scales of the process are both
 perturbative,
as it is the case for $e^{\pm}-onium$ scattering. The application to
$e^{\pm}-p$ scattering requires some assumptions for the coupling to the 
proton. Because of the well-known diffusion in transverse momentum space,
such an application of the dipole model, although 
successful,\cite{nprnprwwal7}
cannot  be considered as a clean test of high-energy perturbative Regge 
dynamics, since non-perturbative effects are expected to be rather 
important and cannot be controled until the sub-leading correction are
 known. A possible test of such a dynamics could be based on single jets 
events
in DIS \cite{jet1} or double jets events in hadron-hadron 
collisions.\cite{mulnav}
Another interesting test of BFKL dynamics would be the $\gamma^*-\gamma^*$
events in $e^+-e^-$ colliders at high energy in the center of mass of the
virtual photon pair and with high (perturbative) photons virtualities.
This has been already proposed in the framework of the original BFKL
equation.\cite{gammagamma}  
Such a process can equivalently be described 
in the dipole picture of BFKL dynamics.\cite{rrw}
 From a phenomenological point of vue, the dipole framework could be applied
to other inclusive processes. The application of this technique for 
exclusive quantities remains however an open question, due to the use
of light-cone quantization, in which the intermediate states are unphysical.

\noindent
\section*{Acknowledgments}

Al this work as been done in collaboration with Henri Navelet.
I wish to thank Jochen Bartels for comments, and  
the Alexander von Humboldt Foundation and 
the II. Instit\"ut f\"ur Theoretische Physik at DESY for support.

\end{document}